\documentclass[%
 aip,
 amsmath,amssymb,
 tightenlines,%
 onecolumn,
]{revtex4-1}

\usepackage{graphicx}% Include figure files
\usepackage{dcolumn}% Align table columns on decimal point
\usepackage{bm}% bold math
\usepackage[utf8]{inputenc}
\usepackage[T1]{fontenc}
\usepackage{mathptmx}
\usepackage{gensymb}

\usepackage[table,xcdraw]{xcolor}
\newcommand{\red}[1]{\textcolor{black}{#1}}

\usepackage{caption}

\begin{document}

\title[Ultra-Slow Acoustic Energy Transport in Dense Fish Aggregates]{Ultra-Slow Acoustic Energy Transport in Dense Fish Aggregates}
% Force line breaks with \\

\author
{Benoit Tallon}
\affiliation{Univ. Grenoble Alpes, CNRS, ISTerre, 38000 Grenoble, France}

\author
{Philippe Roux}
\email[]{To whom correspondence should be addressed; E-mail: philippe.roux@univ-grenoble-alpes.fr}
\affiliation{Univ. Grenoble Alpes, CNRS, ISTerre, 38000 Grenoble, France}

\author
{Guillaume Matte}
\affiliation{iXblue, Sonar division, 13600 la Ciotat, France}

\author
{Jean Guillard}
\affiliation{Univ. Savoie Mont Blanc, INRA, CARRTEL, 74200 Thonon-les-Bains, France}

\author
{John H. Page}
\affiliation{University of Manitoba, Department of Physics \& Astronomy,  Winnipeg, Manitoba R3T 2N2, Canada}

\author
{Sergey E. Skipetrov}
\affiliation{Univ. Grenoble Alpes, CNRS, LPMMC, 38000 Grenoble, France}

\date{\today}% It is always \today, today,

\begin{abstract}
A dramatic slowing down of acoustic wave transport in dense fish shoals is observed in open-sea fish cages. By employing a multi-beam ultrasonic antenna, we observe the coherent backscattering (CBS) phenomenon. We extract key parameters of wave transport such as the transport mean free path and the energy transport velocity of diffusive waves from diffusion theory fits to the  experimental data. The energy transport velocity is found to be about 10 times smaller than the speed of sound in water, a value that is exceptionally low compared with most observations in acoustics.
By studying different models of the fish body, we explain the basic mechanism responsible for the observed very slow transport of ultrasonic waves in dense fish shoals. Our results show that, while the fish swim bladder plays an important role in wave scattering, other organs have to be considered to explain ultra-low energy transport velocities.
\end{abstract}

\flushbottom
\maketitle

\thispagestyle{empty}

\section*{Introduction}

Because of their swim bladder (analogous to an immersed air bubble), bony fish (\textit{Osteichthyes}) are very strong scatterers for underwater acoustic waves.
Thus, (multi-beam) sonar techniques are very efficient to locate and characterize shoals, aggregates or even isolated fish. Most of the fisheries acoustics methods are developed under the single scattering approximation \cite{Simmonds08}, i.e. for low fish concentrations when the wave is scattered at most once during propagation.
For such low densities ($\sim 1$ fish/m$^3$), fish counting is straightforward and efficient using traditional methods such as echo-counting or echo-integration \cite{Foote83}.
However, fish aggregates can be very dense ($\sim 50$ fish/m$^3$) for aquaculture purposes, for example, and in naturally occurring fish schools.
In these cases, the backscattered signal received by the antenna is scattered by several fish during the wave's propagation, which makes fish counting much more challenging.
This multiple scattering regime is problematic for the aquaculture industry, for which nonintrusive biomass estimation is one of the most important issues in regular practice\cite{Li19}.
In a previous work \cite{Tallon20}, we suggested the use of mesoscopic physics and, in particular, of  multiple scattering theory based on the diffusion approximation to deal with wave propagation in dense fish shoals.

In the diffusion approximation, one assumes that after a propagation distance corresponding to the transport mean free path $\ell^*$, the average intensity of multiply scattered waves follows a diffusion process (such as in heat diffusion) with a characteristic diffusivity $D = v_e \ell^*/3$.
The diffusivity involves the energy transport velocity of diffusive waves $v_e = 3D/\ell^*$.  The energy velocity is thus a key parameter describing wave transport, and is %\red{related to } %given by
proportional to 
the ratio of energy flux to energy density.
For diluted or non-resonant systems, $v_e \simeq v_0$ (where $v_0$ is the sound speed in water).
However, as demonstrated with previous model systems, the energy velocity can be highly impacted by resonant phenomena that are typically encountered when the acoustic wavelength becomes similar to the typical size of the scatterers \cite{vanAlbada91,Schriemer97,Tallon17}.

In this paper, we first report an observation of an ultra-low value of the energy transport velocity of diffusive acoustic waves in a dense fish shoal through coherent backscattering (CBS) measurements \cite{vanAlbada85,Wolf85,Tourin97,Akkermans07}.
CBS is a wave interference phenomenon that manifests itself by an enhancement (by a factor of 2) of the average backscattered intensity measured in the direction opposite to that of the incident wave.
The angular profile of backscattered intensity has a cone shape in the stationary (continuous wave) limit, with a width that depends on $\ell^*$.  In the dynamic case, accessed using a pulsed source, the temporal evolution of the backscattering peak's width depends on $D$ and hence on $v_e$.
In order to scan the angular dependence of the backscattered intensity, we employ a multi-beam sonar probe (based on the Seapix technological brick \cite{Mosca16}) in a large fish cage anchored in open sea.
In the second part of the paper, we present a comparative study of energy transport velocity calculations based on Mie theory \cite{Mie08,vanTiggelen92}.
The comparison of four scattering models for individual fish reveals that, while fish are usually approximated as air bubbles in water for scattering of acoustic waves, their complex structure can play an important role for wave transport in dense shoals.
This comparison enables us to identify the essential features that need to be accounted for, and to propose a simple model to replicate the scattering properties of fish in dense shoals.

\section*{Results}

\subsection*{Experiments}

Coherent backscattering (or weak localization) is a mesoscopic phenomenon that has been observed for light \cite{Wolf85,vanAlbada85}, ultrasound \cite{Tourin97}, matter waves \cite{Jendrzejewski12} and seismic waves \cite{Larose04}.
This effect is due to the constructive interference of waves following time-reversed pairs of paths.
This interference produces a peak, centred on the exact backscattering direction, in the angular profile of backscattered intensity. 
In order to observe CBS, we use a Mills Cross multi-beam antenna made of two perpendicular ultrasonic arrays (2$\times$64 transducers with a central wavelength $\lambda =$ 1 cm).
Experiments involve sending a short acoustic pulse (central frequency 150 kHz) into the fish cage by firing all transducers at the same time and performing an angular scan of the backscattered intensity using the beamforming method\cite{Aubry07}.
The cage contains a shoal of gilthead sea breams (\textit{Sparus aurata}) with an average mass of $150$ g and concentration $\eta\sim 50$ fish/m$^3$.

\begin{figure}[ht]
\centering
\includegraphics[width=0.85\linewidth]{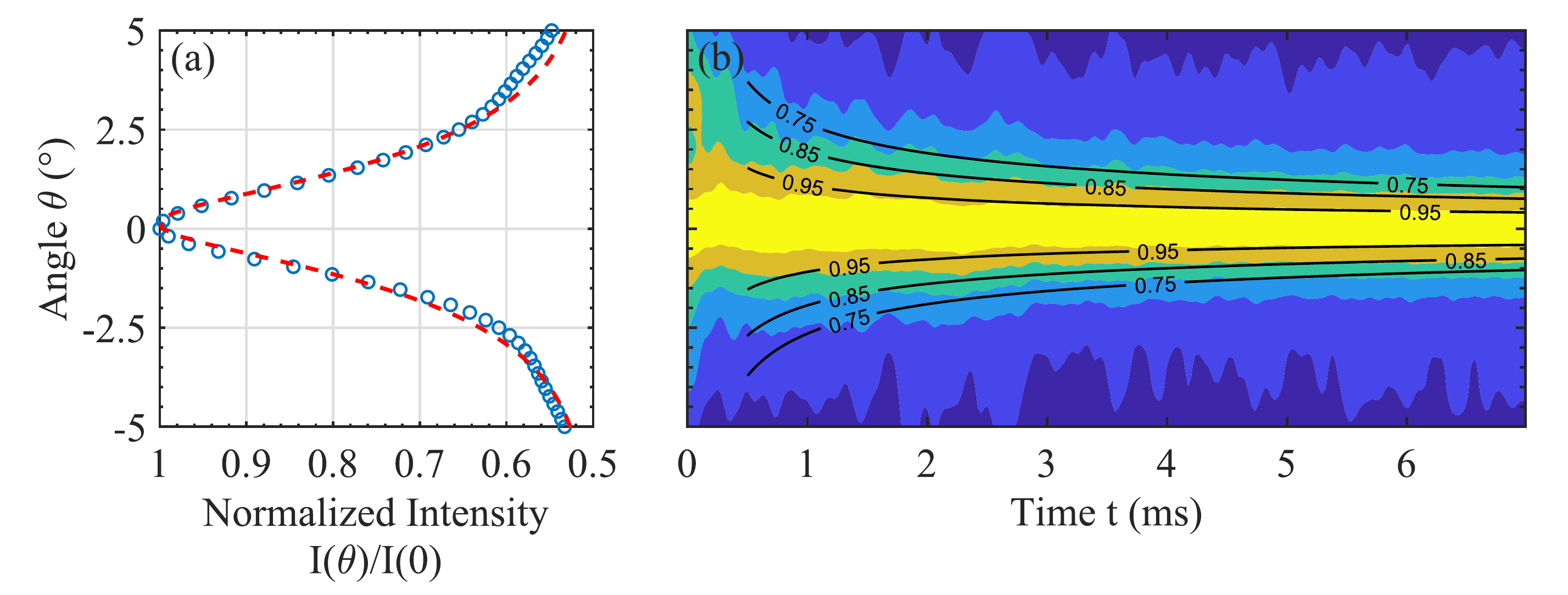}
\caption{(a) Stationary CBS profile (blue open circles) fitted with the diffusion theory (red dashed line). (b) Dynamic CBS profile fitted with the diffusion theory (black solid lines). In (b), the intensity is normalized by its peak value at each time to show more clearly the temporal evolution of the width.} 
\label{f1}
\end{figure}

Time integration of the backscattered intensity yields the \textit{stationary} CBS profile, which has an angular width of $\Delta\theta\sim\lambda/\ell^*$ (where $\lambda$ is the wavelength in water). We fit the entire profile with the predictions of diffusion theory\cite{Akkermans07}; see Fig.~\ref{f1}a.  
The fitting parameters are the transport mean free path $\ell^* =(6.0 \pm 0.2)$ cm and the absorption length $\ell_a =(100 \pm 3)$ cm, which characterizes the exponential decay of acoustic energy due to losses.
The value $k\ell^*= 36\gg1$ (with $k = 2\pi/\lambda$) indicates that even if sound is strongly scattered in this shoal, no complex phenomena such as strong localization (occurring for $k\ell^*\sim1$) impact the CBS peak shape.

The time-resolved \textit{dynamic} CBS peak narrows with time (Fig.~\ref{f1}b).
From diffusion theory\cite{Akkermans07}, the CBS width depends on diffusivity as $\Delta\theta\propto1/\sqrt{Dt}$.
Using the measured value of $\ell^*$ and the diffusion theory, the fitting of the dynamic CBS peak gives us $D = (1.3\pm0.1)$ m$^2$/s.
In this way, the simultaneous measurement of $\ell^*$ and $D$ leads to the energy velocity $v_e = 3D/\ell^* = (65\pm5)$ m/s.
This value of $v_e$ is surprising low (an order of magnitude lower than the sound speed in water $v_0 = $ 1500 m/s) , which is a very rare observation for acoustic waves\cite{Schriemer97,Viard15,Tallon17,Tallon20_1,Tallon20}.
Similarly slow diffusion has been observed previously with CBS measurements in dense shoals of sea breams, sea basses (\textit{Dicentrarchus labrax}) or croakers (\textit{Argyrosomus regius}) \cite{Tallon20}, but no such extreme behaviour has been found for acoustic waves in other multiply scattering media\cite{Schriemer97,Viard15,Tallon17,Tallon20_1}.
However, unlike this past work on dense fish shoals, the cage considered in this paper has a sufficiently low fish concentration to allow us to neglect mesoscopic interferences that might impact the energy velocity, thereby facilitating a quantitative interpretation of the current results.
In the following section, we employ a model \cite{vanTiggelen92} that takes into account the scattering delay induced by the fish and explains this ultra low value of energy velocity.

\subsection*{Energy velocity calculation}

Here, we employ a microscopic model of the energy velocity\cite{vanTiggelen92} in order to identify the mechanisms responsible for the ultra low $v_e$ value.
This description is based on a model that accounts for the delay $\Delta t_{ave}$ induced by an immersed scatterer, and predicts a result for $v_e$ that may be approximated as \cite{Tallon20_1}
\begin{equation}\label{eq1}
\frac{1}{v_e} \approx \frac{1}{v_{gr}} + \eta\sigma_T\Delta t_{ave}.
\end{equation}
Here $v_{gr}$ is the group velocity of the average wave field ($v_{gr} \sim v_0$ in the absence of dispersion effects) and $\eta$ is the scatterer concentration. $\Delta t_{ave}$ is the scattering delay time for a single scattering event, calculated from the intensity-weighted angle-averaged phase derivative with frequency of the scattering amplitude [equation~(\ref{eqA4})], and $\sigma_T$ is the total scattering cross section.
Thus, to obtain slow diffusive waves, the system has to be dense (high $\eta$) with strong scatterers (high $\sigma_T$) and large scattering delay (large $\Delta t_{ave}$).  

\begin{figure}[ht]
\centering
\includegraphics[width=0.9\linewidth]{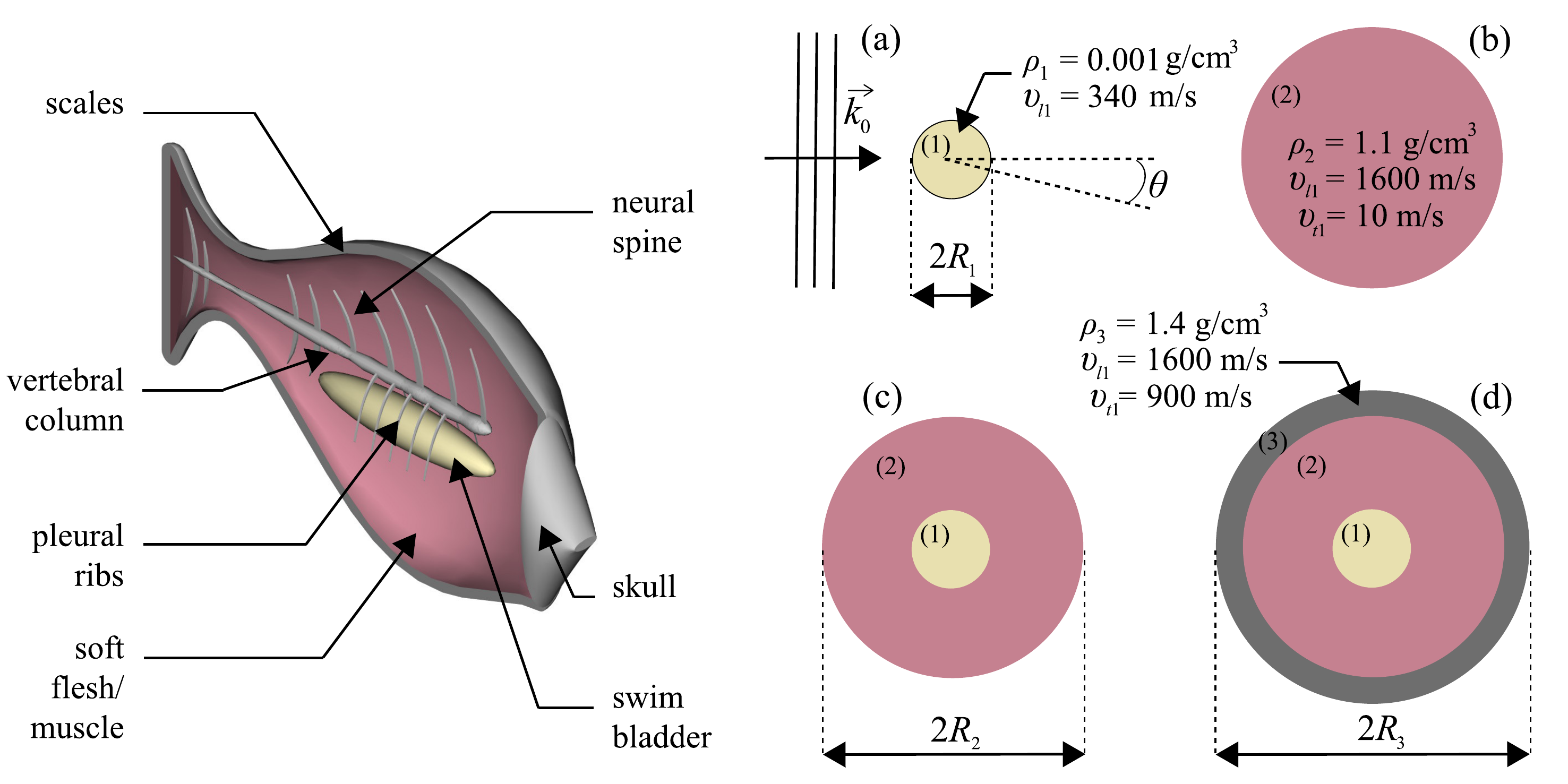}
\caption{On the left: Schematic representation of a sea bream body. On the right: Models a, b, c and d used for the energy velocity calculation.}
\label{f2}
\end{figure}

In order to identify the parts of the fish that are responsible for the ultra-low energy velocity, we use equation (\ref{eq1}) to compute $v_e$ for four idealized spherically symmetric scattering models representing different simplifications of the complex fish body (Fig. \ref{f2}). Since this theory enables analytic expressions to be obtained for relatively simple scattering geometries, the goal here is to capitalise on this capability to search for the key features that need to be considered in order to understand the origin of the remarkably slow energy velocity.  Thus, rather than attempting a complex simulation that might obscure the basic scientific mechanism(s) at play, we focus on very simple models to reveal the basic scattering mechanisms involved:
\begin{itemize}
    \item Model a: an air bubble representing the fish swim bladder with radius $R_1 =$ 10 mm (Fig. \ref{f2}a), longitudinal wave speed $v_{l1} = 340$ m/s and density $\rho_1 =$ 0.001 g/cm$^3$.
    \item Model b: a homogeneous soft sphere representing the fish flesh with radius $R_1 =$ 76 mm, longitudinal wave speed $v_{l2} = 1600$ m/s, shear wave speed $v_{t2} = 10$ m/s and density $\rho_2 =$ 1.1 g/cm$^3$ (Fig. \ref{f2}b).
    \item Model c: a combination of models a and b representing the swim bladder surrounded by a flesh layer (Fig. \ref{f2}c).
    \item Model d: similar to model c with an additional hard thin layer representing fish scales and bones with radius $R_3 =$ 78 mm, longitudinal wave speed $v_{l3} = 1600$ m/s, shear wave speed $v_{t3} = 900$ m/s and density $\rho_3 =$ 1.4 g/cm$^3$ (Fig. \ref{f2}d).
\end{itemize}
Some scattering theories allow calculations for spheroids\cite{Flammer14} that might be closer to the actual fish shape, but none of those calculations are developed for multilayer scatterers.
However, the spherical approximation is suitable in the present case because of the randomized fish orientation in the azimuthal plane.
Thus on average, the effective scatterer shape seen by the incident plane wave can be approximated as a sphere.

\begin{figure}[ht]
\centering
\includegraphics[width=1\linewidth]{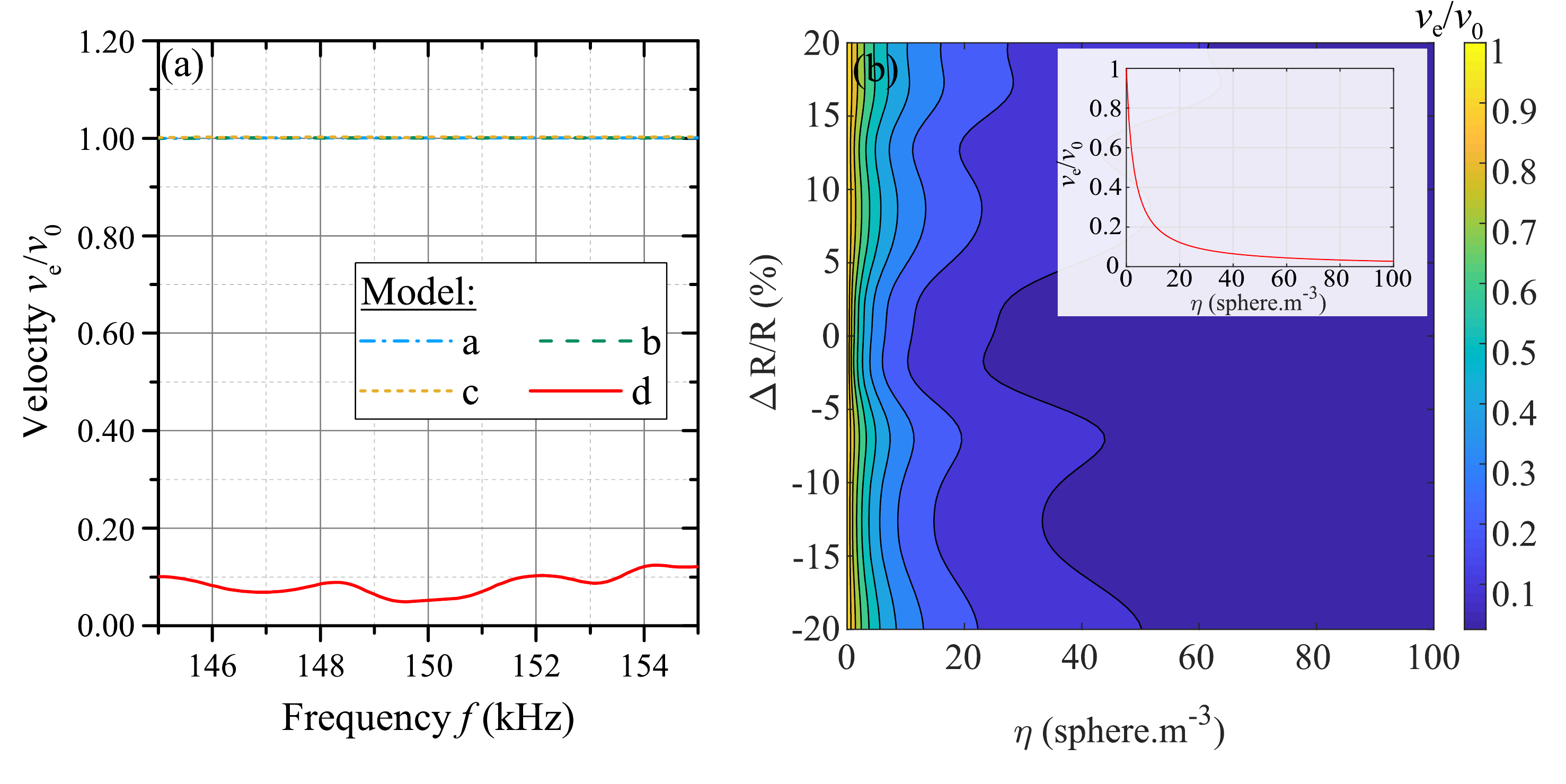}
\caption{(a) Normalized velocity $v_e/v_0$ calculations versus frequency for the four different models. (b) Normalized velocity $v_e/v_0$ calculations versus fish size variation $\Delta R/R$ and concentration $\eta$ at a given frequency $f =$ 150 kHz for model d. The inset represents $v_e/v_0$ versus fish concentration for $\Delta R/R$ = 0 and $f =$ 150 kHz.}
\label{f3}
\end{figure}

Model a represents the usual assumption in fisheries acoustics (in the single scattering regime): since, at least near resonance, the swimbladder is the most reflective organ for acoustic waves, fish shoals are often seen as clouds of air bubbles in water.
However, as shown in Fig. \ref{f3}a, Eq. (\ref{eq1}) applied to model a fails to explain the ultra-low value of energy velocity found in our experiments.
The same conclusion can be drawn for models b and c, for which we also obtain $v_e\sim v_0$.
On the other hand, model d predicts a very slow energy velocity, $v_e \approx$ 100 m/s, at the frequency $f =$ 150 kHz, and ranges from about 50 to 150 m/s over the bandwidth of the transducers.
In comparison with the other models, these values are very close to our measurement ($v_e = 65\pm5$ m/s).
The phase and group velocities (see Supplementary Information, Fig.~S2) can also be calculated for these models.
Both of these velocities are found to be very close to $v_0$ for all four models.
These results indicate that the complex fish structure mostly impacts only the diffusive waves; the ballistic wave velocities (the ``line-of-sight'' propagation through the shoal) are not significantly affected by the fish scattering and concentration.

Thus, the hard thin layer surrounding the soft solid representing fish flesh seems to play an important role for the slowing down of diffusive acoustic waves.
Figure \ref{f3}(b) shows the calculation of energy velocity for a range of fish concentrations $\eta$ and as a function of size variation $\Delta R/R$ (here the size variation $\Delta R/R$ is a factor that is equally applied to the three radii $R_1$, $R_2$ and $R_3$).
It is important to note the weak size dependence of energy velocity on $\Delta R/R$ that proves that the drop of $v_e/v_0$ in model d with respect to model c is not due to size differences (outer radius $R_3$ versus $R_2$).
In contrast, the fish concentration $\eta$ seems to have a significant impact on $v_e$.
This effect could be interesting for enabling a new method of biomass assessment with acoustic waves compared with more traditional acoustic methods (see, for example, Ref. \cite{Foote83}).

\section*{Discussion}

In this section, we interpret the role of each part of the fish model on the scattering delay.
The double core-shell structure of model d seems to explain the observed ultra-low energy velocity.
As has been observed in the past \cite{Liu00}, core-shell scatterers can indeed exhibit very strong scattering.
However, no slow diffusive waves have been measured for such scatterers in the past.
Resonant mechanisms have been identified \cite{vanAlbada91,Schriemer97,Tallon17} as being responsible for the decrease of $v_e$ in both optics and acoustics, but only for homogeneous scatterers.
These effects result in large energy velocity variations, with $v_e \sim v_0$ far from resonant frequencies and $v_e \ll v_0$ around resonances.
In the present case, such frequency fluctuations (expected over several tens of kHz) cannot be observed in our experiments, since the sonar bandwidth is too narrow; hence, frequency-resolved measurements were not feasible, and our experimental determination of $v_e$ corresponds to a narrow bandwidth-limited average around the central frequency (150 kHz) of the transducers.  
%(in order not to compromise good time resolution in measuring the evolution of the dynamic CBS peak) c
Thus, from experimental observations, we can measure conclusively the low $v_e$ value but we cannot conclude anything about its potential variations over a larger frequency range.

\begin{figure}[ht]
\centering
\includegraphics[width=0.85\linewidth]{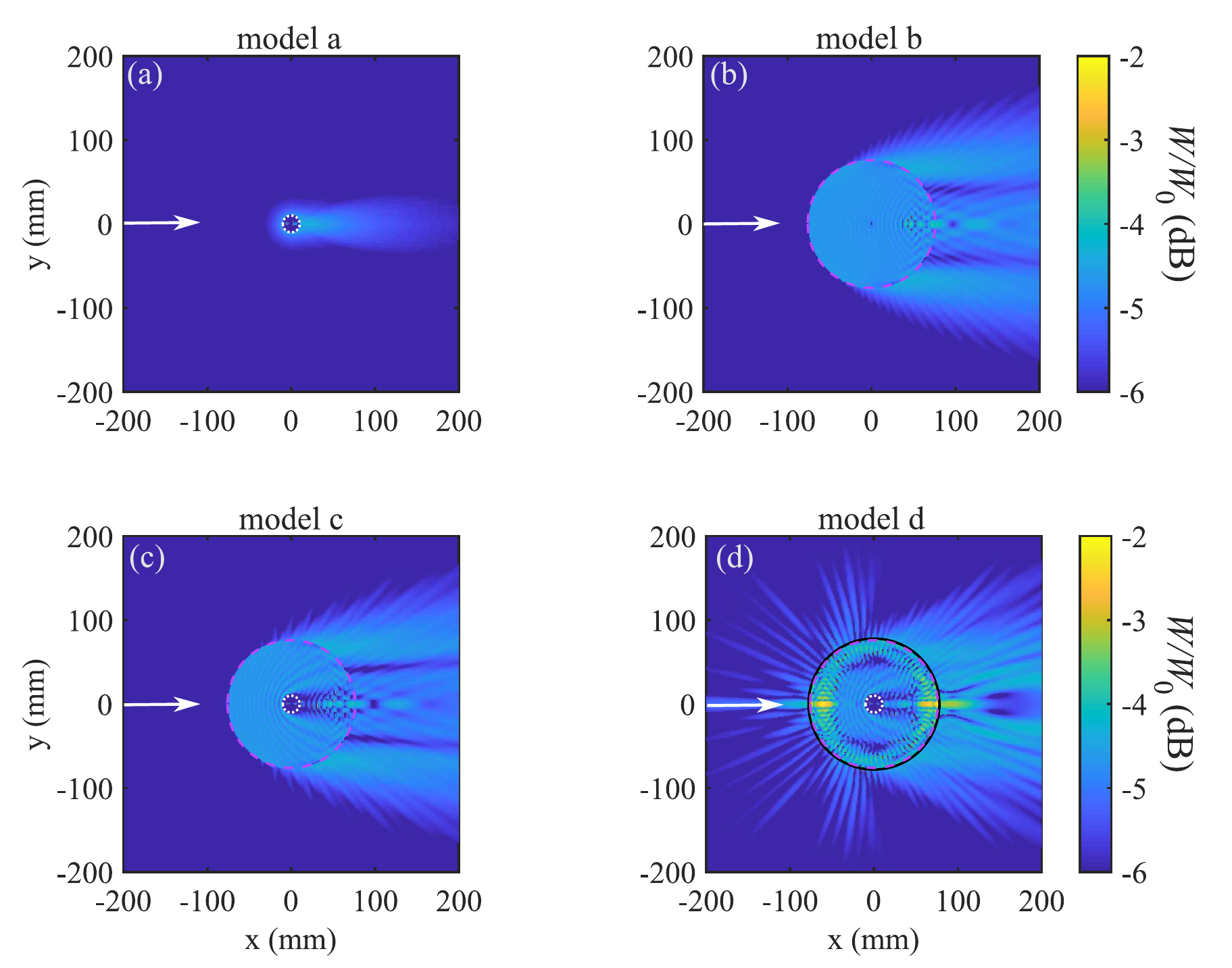}
\caption{Calculations of the longitudinal acoustic energy density $W$ for the four different model scatterers at $f=150$ kHz. $W_0$ is the incident energy density. The white dotted line represents the limits of the swim bladder, the dashed pink line the flesh layer and the solid black line the thin scales and bones layer. White arrows indicates the direction of the incident plane wave.}
\label{f4}
\end{figure}

A way to interpret the scattering delay impacting the energy velocity is the calculation of the acoustic energy density inside and outside the scatterer\cite{Cowan98,Tallon20_1}.
High values of energy density suggest that waves are ``stored'' in the scatterer.
This energy is then released in the surrounding medium with a certain delay resulting in a slowing down of diffusive wave transport.
Figure \ref{f4} shows the energy density calculations for longitudinal waves at the frequency $f =$ 150 kHz for the four different model scatterers.

The energy density calculation for an air bubble (model a) exhibits predominantly  forward scattering as expected at high frequency for small scatterers with large acoustic contrast\cite{Hulst81}.
Since the frequency being considered in this study is far from the resonance of the bubble ($f_{res} \sim$ 0.5 kHz), the scattering is very weak for this model, and the energy density inside the scatterer is small.
Thus, one might expect only a modest scattering delay, and the scattering strength (or total cross section $\sigma_T$) induced by the isolated swim bladder is not high enough for the product of scattering delay and cross section to cause a significant decrease of $v_e$.
For models b and c, both the energy density inside the scatterers and the forward scattering are somewhat larger than for model a, suggesting that there could be a bigger difference between group and energy velocities, but additional information would be needed to assess if they could be interesting candidates for predicting a slow energy velocity.
However, for the double core-shell system (model d), the result of the energy density calculation is much more striking, as we observe a \emph{very} strong increase of scattered energy density (the large increase in wave energy stored in the fish body is due to the hard scales and bones layer). The subwavelength-scale outer layer helps the generation of slow shear waves via mode conversion and the trapping of both longitudinal and shear waves in the fish flesh.
This stored acoustic energy is then re-radiated into the surrounding water with a large delay\cite{vanTiggelen92}.

\begin{figure}[ht]
\centering
\includegraphics[width=1\linewidth]{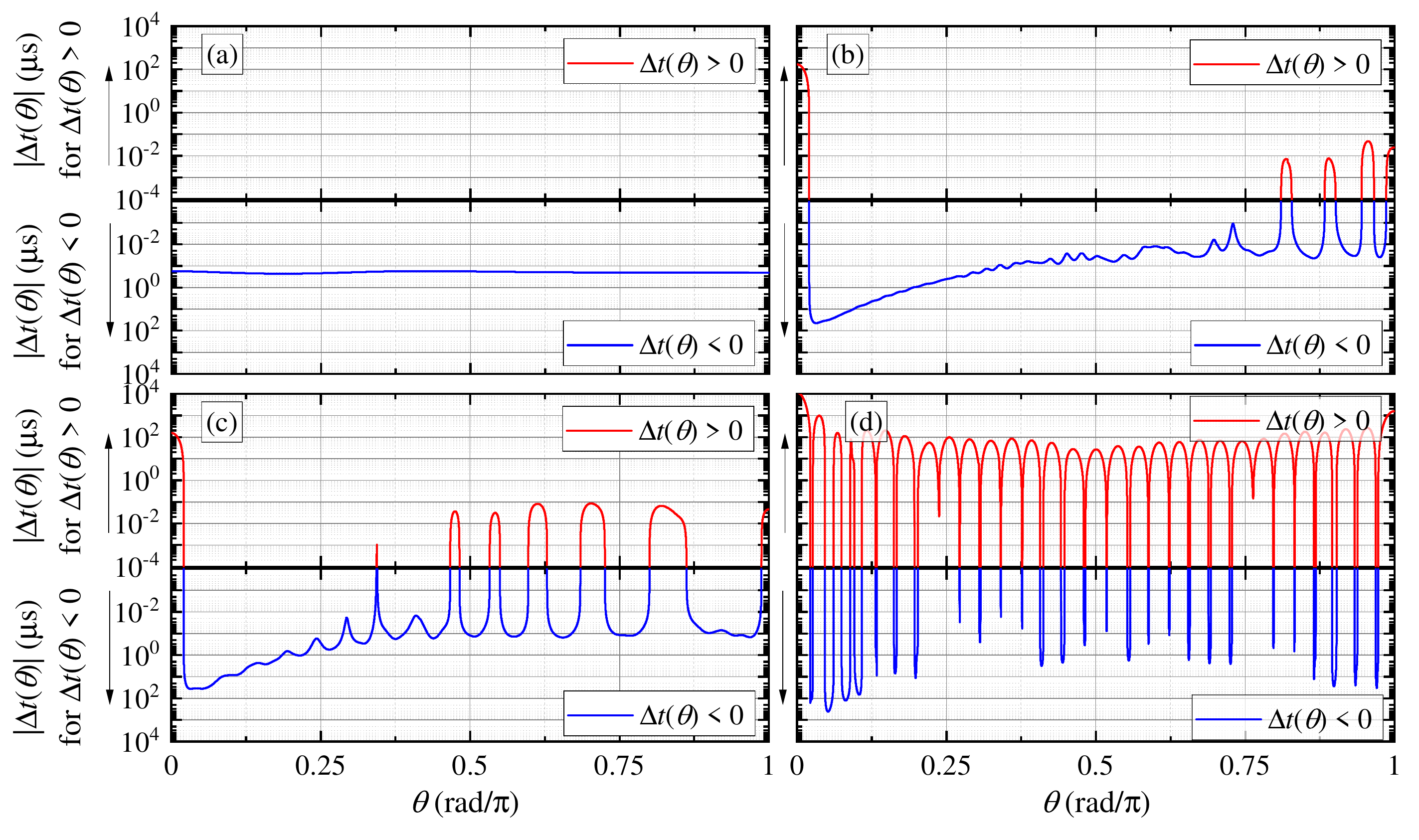}
\caption{(a) to (d) Angular dependence of the normalized scattering delays $\Delta t_{\theta}$ calculated for the four models (a, b, c and d), at a frequency of 150 kHz.}
\label{f5}
\end{figure}

The link between the stored energy and the large scattering delay becomes clear when comparing Fig. \ref{f4} and Fig. \ref{f5}, which shows the angular dependence of the scattering delay $\Delta t(\theta)$ %normalized by the wave period $T_0$, 
for all models [Fig.~\ref{f5}(a)-(d)]  (see the Supplementary Information for calculation details).  In particular, this figure shows that the angle-resolved scattering delays are much larger, typically by a couple of orders of magnitude, for model d than for the other models, for which the energy densities inside the scatterers are much less. Specifically, for model a, the delay is relatively small and negative at all scattering angles, %and positive (indicating a slight decrease in the energy velocity relative to the group velocity) for all angles, 
whereas for models b and c, the delays are negative for most angles and have, typically, somewhat larger magnitudes.  In all three cases, these results suggest that the total angle averaged delay will be fairly small and certainly negative (indicating a slight enhancement of the energy velocity relative to the group velocity).  For model d, however, very large delays, both positive and negative, are seen, with the positive delays dominating.  These observations are confirmed by doing the angular integration of $\Delta t(\theta)$, yielding average scattering delays per wave period $\Delta t_\text{ave}/T_0$ for models a through d of $-0.4, -1.4, -1.4$ and $80.8$, respectively. Thus, we find that adding the hard coating increases the magnitude of the angle-averaged delay by a factor of approximately 100 or more in comparison with the other models.

\begin{figure}[ht]
\centering
\includegraphics[width=0.85\linewidth]{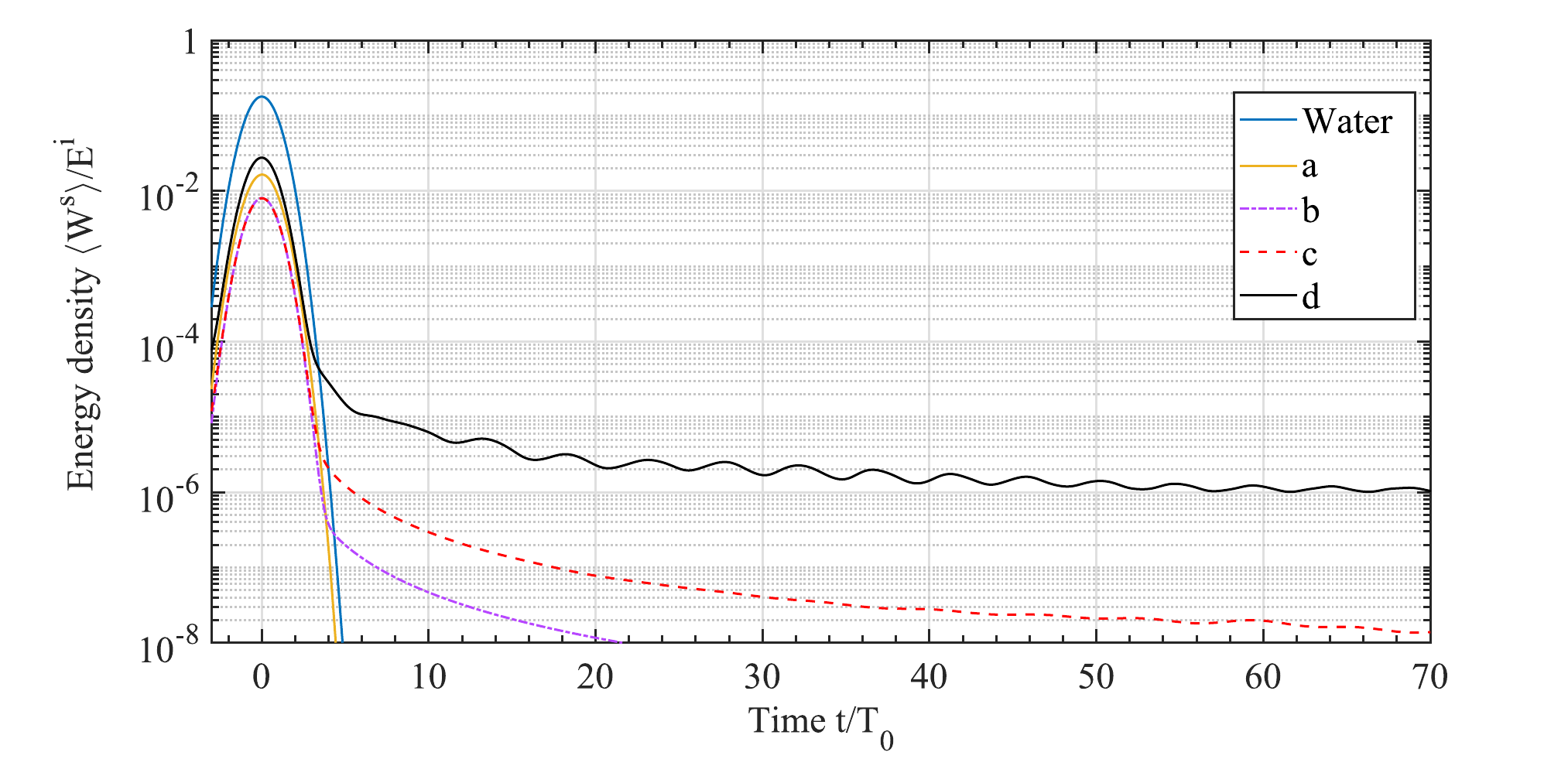}
\caption{Temporal evolution of scattered energy density $\langle W^s \rangle$  (plots are normalized by the total incident energy density $E^i$). The solid blue line represents the energy density of incident pulse (without a scatterer). The time axis is normalized by the width of the incident pulse $T_0 =$ 0.1 ms.
\label{f6}}
\end{figure}

To further illustrate this large scattering delay, we calculate the dynamic energy density of the scattered waves $\langle W^s\rangle$ outside the scatterers (Fig.~\ref{f6}).
$\langle W^s\rangle$ is obtained by integrating $W(f)$ (c.f., Fig.~\ref{f4} at $f=150$ kHz) over the bandwidth of the incident pulse.
The resulting temporal evolution of $W$ is spatially averaged on the region $R_3<r<b$ (where $r = \sqrt{x^2 + y^2}$ and $b =$ 85 mm is the average inter-scatterer distance for the  concentration $\eta$  considered here.
In this way, we obtain the dynamics of the acoustic energy density in the surroundings of each scatterer.

The higher overall energy obtained with model d confirms the strong influence on the scattering induced by the presence of scales and bones.
Furthermore, while models a, b and c all predict that $\langle W_d \rangle$ decreases quickly (which explains $v_e\sim v_0$ for these systems), a much slower decay of $\langle W_d \rangle$ with time is obtained with model d.
%Furthermore, while models a, b and c exhibit quick decrease of $\langle W_d \rangle$ (which explains $v_e\sim v_0$ for these systems), a very slow exponential decay of $\langle W_d \rangle$ with time is obtained with model d.
This slow decay for model d demonstrates clearly that the {large} stored acoustic energy in the fish body (Fig.~\ref{f4}(d)) is slowly radiated into the surrounding medium (Fig.~\ref{f6}).
To summarize, the core-shell model system with swim bladder + flesh + scales-and-bones layer leads to a large slowly decaying energy density associated with a large scattering delay.
These observations and the quite large fish concentration $\eta$ successfully explain the slow diffusion of acoustic waves observed in the sea bream shoal.

\section*{Conclusion}

In conclusion, while the assumption that fish shoals are equivalent to air bubble clouds is adequate for most acoustic studies\cite{Simmonds08}, some cases with dense fish shoals have to be considered more carefully.
Indeed, high fish concentration ($\eta\sim$ 50--100 fish/m$^3$) can lead to an accumulation of scattered energy, a resulting increase in scattering delay and a drastic slowing down of sound diffusion.
Thus, it is essential to properly estimate $v_e$ in fish shoals in order to extend the range of application of acoustic fisheries techniques to very dense shoals.
Several improvements (such as scatterer shape and structure) can, in principle, be implemented in the model in order to calculate more accurate $v_e$ values.
Nonetheless, as a first study, the simple models used here are sufficient to reveal the essential wave  physics behind our observations, and enable us to demonstrate the strong impact of coating layers surrounding the swim bladder on wave transport in dense shoals.

As shown in Fig. \ref{f3}b, the strong energy velocity dependence on the fish concentration could be useful for fish counting purposes.
The method is particularly interesting since $v_e$ depends strongly on concentration and only weakly on fish size variation (over a reasonable size range for aquaculture conditions: $\Delta R/R\leqslant\pm$ 15\%).
In particular, such a tool could help the aquaculture industry for which large fish concentrations make acoustic biomass estimation impossible using traditional acoustic approaches based on the assumption of single scattering.
Noninvasive monitoring of fish farms using diffusive waves\cite{Tallon20,Tallon20_2} is currently under study with long-term experiments (several weeks) in order to investigate the impact of fish size and biomass variations on the diffusive transport of sound waves.

\section*{Methods}

The sea cages are \red{located} in the Mediterranean Sea (Cannes, France) where the water temperature is about 20\degree and salinity 3.6\%. The cage in which experiments were conducted is cubic with a volume of 125 m$^3$.
The distance from the bottom of the cage to the sea bottom is $z =$ 6.5 m.
To maintain the organic label of the farm and to avoid the need for drug treatments, the fish densities in these cages are lower than those in intensive farming facilities (where mass densities can reach 100 kg/m$^3$).
The feeding procedures are controlled to obtain a calibrated fish size.

The theory used to fit the experimental CBS data is derived from a diffusion equation for the average intensity $\langle I(\mathbf{r},t) \rangle$ \cite{Akkermans07} that is solved for a disordered medium occupying the half-space $z > 0$, with a delta-function source at $z = z'=\ell^*$ \cite{Carslaw59}:
\begin{equation}\label{eqS1}
\langle I(t) \rangle = \frac{I_0}{2\pi}\int_{-\infty}^{+\infty}\frac{z_0\textrm{exp}(-\gamma_0 z')}{D(1+\gamma_0 z_0)}\textrm{exp}(-\textrm{i}\Omega t)d\Omega,
\end{equation}
where $\gamma_0^2(\Omega) = \frac{-\textrm{i}\Omega}{D} + \frac{1}{D\tau_a}$, $\tau_a$ is the characteristic absorption time and $z_0 = \frac{2}{3}\frac{1+R}{1-R}\ell^*$ is the extrapolation length, with $R =$ 0.99 as the reflection coefficient for the water/air interface.  Theoretical predictions for both the dynamic and stationary CBS peaks can then be obtained using this theory\cite{Tallon20}.%, as outlined in the Supplementary Information and described more fully in Ref. 4.

The energy velocity theory\cite{vanAlbada91} (Eq. \ref{eq1}) is based on the calculation of \red{the} scattering function $F(\theta) = |F(\theta)|\textrm{e}^{\textrm{i}\varphi(\theta)}$ with magnitude $|F(\theta)|$ and phase $\varphi(\theta)$. $F(\theta)$ represents the scattering amplitude in the direction given by the angle $\theta$ with respect to the incident wavevector, and is given by the following expression:
\begin{equation}\label{eqA3}
F(\theta) = \frac{1}{ik_\textrm{0}}\sum_\textrm{n}(2\textrm{n} + 1)A_\textrm{n}P_\textrm{n}(\cos\theta),
\end{equation}
where $k_0$ represents the wavenumber of incident wave, $P_n$ the Legendre polynomials and $A_n$ the scattering amplitude coefficients of the scattered field.
\red{The} $A_n$ coefficients are obtained by solving the Mie problem\cite{Mie08}\red{, invoking} the calculation of stress and displacement continuity conditions (for both \red{longitudinal} and shear waves) at the boundaries $R_1$, $R_2$ and $R_3$ \cite{Faran51}.
$F(\theta)$ is then used to obtain the scattering cross-section $\sigma$ and delay $\Delta t_{ave}$ that are needed for \red{the} calculation of \red{equation (\ref{eq1})}:
\begin{equation}\label{eqA4}
\sigma = \red{2\pi} \int{d\theta \sin \theta \left| f(\theta) \right|^2}\ \hspace{1cm} \textrm{and}\ \hspace{1cm}
\Delta t_{ave} = \frac{\int{d\theta \sin \theta \left| f( \theta ) \right|^2}\frac{\partial \varphi} {\partial \omega} }{\int{d\theta \sin \theta \left| f( \theta) \right|^2}}.
\end{equation}
Additional details on the experimental method and theoretical calculations can be found in the Supplementary Information.

\section*{Author contributions statement}

B.T., P.R. and G.M. conducted the experiment. B.T., J.H.P., S.E.S. and P.R. developed the models. All authors analysed the results and reviewed the manuscript.

\section*{Acknowledgments}

The authors wish to thank Olivier Lerda from iXblue for Seapix signal processing routines. JHP is grateful for support from the Natural Sciences and Engineering Research Council of Canada's Discovery Grant Program (RGPIN-2016-06042).

\end{document}

\title[Ultra-Slow Acoustic Energy Transport in Dense Fish Aggregates]{Supplementary Information for \\ ``Ultra-Slow Acoustic Energy Transport in Dense Fish Aggregates"}
% Force line breaks with \\

\author
{Benoit Tallon}
\affiliation{Univ. Grenoble Alpes, CNRS, ISTerre, 38000 Grenoble, France}

\author
{Philippe Roux}
\email[]{To whom correspondence should be addressed; E-mail: philippe.roux@univ-grenoble-alpes.fr}
\affiliation{Univ. Grenoble Alpes, CNRS, ISTerre, 38000 Grenoble, France}

\author
{Guillaume Matte}
\affiliation{iXblue, Sonar division, 13600 la Ciotat, France}

\author
{Jean Guillard}
\affiliation{Univ. Savoie Mont Blanc, INRA, CARRTEL, 74200 Thonon-les-Bains, France}

\author
{John H. Page}
\affiliation{University of Manitoba, Department of Physics \& Astronomy,  Winnipeg, Manitoba R3T 2N2, Canada}

\author
{Sergey E. Skipetrov}
\affiliation{Univ. Grenoble Alpes, CNRS, LPMMC, 38000 Grenoble, France}

\date{\today}% It is always \today, today,

\begin{abstract}
\end{abstract}

\maketitle

\section*{Introduction}
This document provides additional information on the experiments that were performed to investigate dense fish shoals in open-sea fish cages, and on the theory used to interpret the results of these measurements.

\section*{Experimental set-up}

%The sea cages are \red{located} in the Mediterranean Sea (Cannes, France) where the water temperature is about 20\degree and salinity 36\textperthousand. \blue{[COMMENT: In view of Sergey's remark, I think it would be clearer to indicate the salinity in percent, as $3.6\%$.]}
%The cage in which experiments were conducted is cubic with a volume of 125 m$^3$.
%The distance from the cage end to the sea bottom is $z =$ 6.5 m.
%To maintain the organic label of the farm and to avoid the need for drug treatments, the fish densities in these cages are lower than those in intensive farming facilities (where mass densities can reach 100 kg/m$^3$).
%The feeding procedures are controlled to obtain a calibrated fish size.

The CBS experiments were performed using %acoustic device used is 
a Mills cross-shaped antenna that consists of two perpendicular linear arrays, each with 64 ultrasonic piezoelectric transducers.
Each transducer is narrow band (bandwidth, 10 kHz) with a central frequency of 150 kHz, which corresponds to a wavelength $\lambda \simeq$ of 1 cm in water. The transducers are square in shape and have sides of 0.5 cm $\simeq \lambda$/2.
The antenna (expressly designed by the iXblue company for aquaculture monitoring) is placed just below the water surface, facing the sea bottom.

At full power, the radiation pressure of the ultrasonic antenna is 133 \red{dB$\cdot \mu$Pa} (measured at a distance \red{$d=1$} m from the source and for the frequency \red{$f=150$} kHz). The hearing threshold for sciaenidae fish is about 60 \red{dB$\cdot \mu$Pa} in the frequency range $f\in[0.1-1]$ kHz \cite{ladich2013}. No avoidance behavior of the fish \red{has} been observed during our experiments.
For comparison, some vessel noise signatures are reported in \red{Ref. S2} %\cite{simmonds08} 
and turn out to be around 130 \red{dB$\cdot \mu$Pa} in the hearing frequency range of sciaenidae fish.

\section*{Diffusion theory}

As indicated in the Methods section of the main paper, the theory used to interpret the experimental CBS data is obtained from a diffusion model for the average intensity\cite{Akkermans07} $\langle I(\mathbf{r},t) \rangle$ %that is solved 
in a disordered medium occupying the half-space $z > 0$ when the source is a delta-function  at $z = z'=\ell^*$\cite{Carslaw59}.  The expression for $\langle I(\mathbf{r},t) \rangle$ is repeated here for convenience: 
\begin{equation}\label{eqA1}
\langle I(t) \rangle = \frac{I_0}{2\pi}\int_{-\infty}^{+\infty}\frac{z_0\textrm{exp}(-\gamma_0 z')}{D(1+\gamma_0 z_0)}\textrm{exp}(-\textrm{i}\Omega t)d\Omega,
\end{equation}
where $\gamma_0^2(\Omega) = \frac{-\textrm{i}\Omega}{D} + \frac{1}{D\tau_a}$, $\tau_a$ is the characteristic absorption time and $z_0 = \frac{2}{3}\frac{1+R}{1-R}\ell^*$ is the extrapolation length, with $R =$ 0.99 as the reflection coefficient for the water/air interface.
The theoretical expression used to fit the dynamic CBS profile $\langle I(\theta, t) \rangle $ follows from the same diffusion theory, and is given by %as was used to obtain equation (\ref{eqA1}) for the mean intensity:} %the expression for the mean intensity (\ref{eqA1}):
\begin{equation}\label{eqA2}
\langle I(\theta, t) \rangle = \frac{I_0}{2\pi}\int_{-\infty}^{+\infty}
\frac{z_0}{D}
\left \{\frac{\textrm{e}^{-\gamma_0z'}}{1 + \gamma_0z_0} + \frac{\textrm{e}^{-\gamma z'}}{1 + \gamma z_0}\right \}\textrm{exp}(-\textrm{i}\Omega t)d\Omega,
\end{equation}
where $\gamma^2(\theta,\Omega) = \frac{-\textrm{i}\Omega}{D} + k_0^2\textrm{sin}^2(\theta) + \frac{1}{D\tau_a}$ and $\gamma_0 = \gamma(\theta = 0,\Omega)$.  The expression for the stationary CBS profile is obtained by integrating equation (\ref{eqA2}) over time\cite{Tallon20b}.  

\section*{Phase and group velocities}

We \red{outline} %develop 
the calculation of the wave number $k$ for the average field.
This wave number, useful for the energy velocity calculation, is expressed as :
\begin{equation}\label{eqA3}
    k^2 = k_0^2 + 4\pi\eta F(\theta = 0),
\end{equation}
where $\eta$ is the scatterer concentration and $k_0 = \omega/v_0$ is the wave number of the incident wave at frequency $f = \omega/2\pi$ in water. In the far field approximation, the scattering function $F(\theta) = |F(\theta)|\textrm{exp}[i\varphi(\theta)]$, with magnitude $|F(\theta)|$ and phase $\varphi(\theta)$, represents the scattering amplitude in the direction given by the angle $\theta$ with respect to the incident wave vector. 
$F(\theta)$ is given by the following expression 
\begin{equation}\label{eqA4}
F(\theta) = \frac{1}{ik_\textrm{0}}\sum_\textrm{n}(2\textrm{n} + 1)A_\textrm{n}P_\textrm{n}(\cos\theta),
\end{equation}
where $P_n$ represents the Legendre polynomials and $A_n$ the scattering amplitude coefficients of the scattered field. \red{The} $A_n$ coefficients are obtained by solving the Mie problem\cite{Mie08}: by imposing stress and displacement continuity conditions at the interfaces of a spherical scatterer (model a), one obtains a system of 3 linear equations with 3 coefficients\cite{Faran51}. %a the calculation of the $3\times 3$ \red{system} of linear equations given by  \cite{Faran51}.
The \red{solution of this system of equations} requires inverting a $3\times 3$ matrix, and leads to all of the 3 partial wave amplitudes (the scattered longitudinal wave, the refracted longitudinal wave and the refracted shear wave).
The calculation of scattering amplitude of a coated sphere (model c) is obtained by solving a system of 7 linear equations with 7 coefficients, as %$7\times 7$ system of linear equations
detailed in Ref. \red{S8}. %$\cite{Jing92}.
Following the same procedure, the last case of a double coated sphere (model d) \red{requires the solution} of a system of 10 linear equations with 10 coefficients %the $10\times 10$ system of linear equations 
given by stress and displacement continuity conditions at the 3 interfaces.
%the calculation of the $3\times 3$ \red{system} of linear equations given by stress and displacement continuity conditions at the interfaces of the spherical scatterer \cite{Faran51}. The \red{solution of this system of equations} leads to all of the 3 partial wave amplitudes (the scattered longitudinal wave, the refracted longitudinal wave and the refracted shear wave). The calculation of scattering amplitude of a coated sphere (model c) is obtained by solving the $7\times 7$ system of linear equations detailed in Ref. \red{S8}. %$\cite{Jing92}. Following the same procedure, the last case of a double coated sphere (model d) \red{requires the solution} of the $10\times 10$ system of linear equations given by stress and displacement continuity conditions at the 3 interfaces.

\begin{figure}[ht]
\centering
\includegraphics[width=0.8\linewidth]{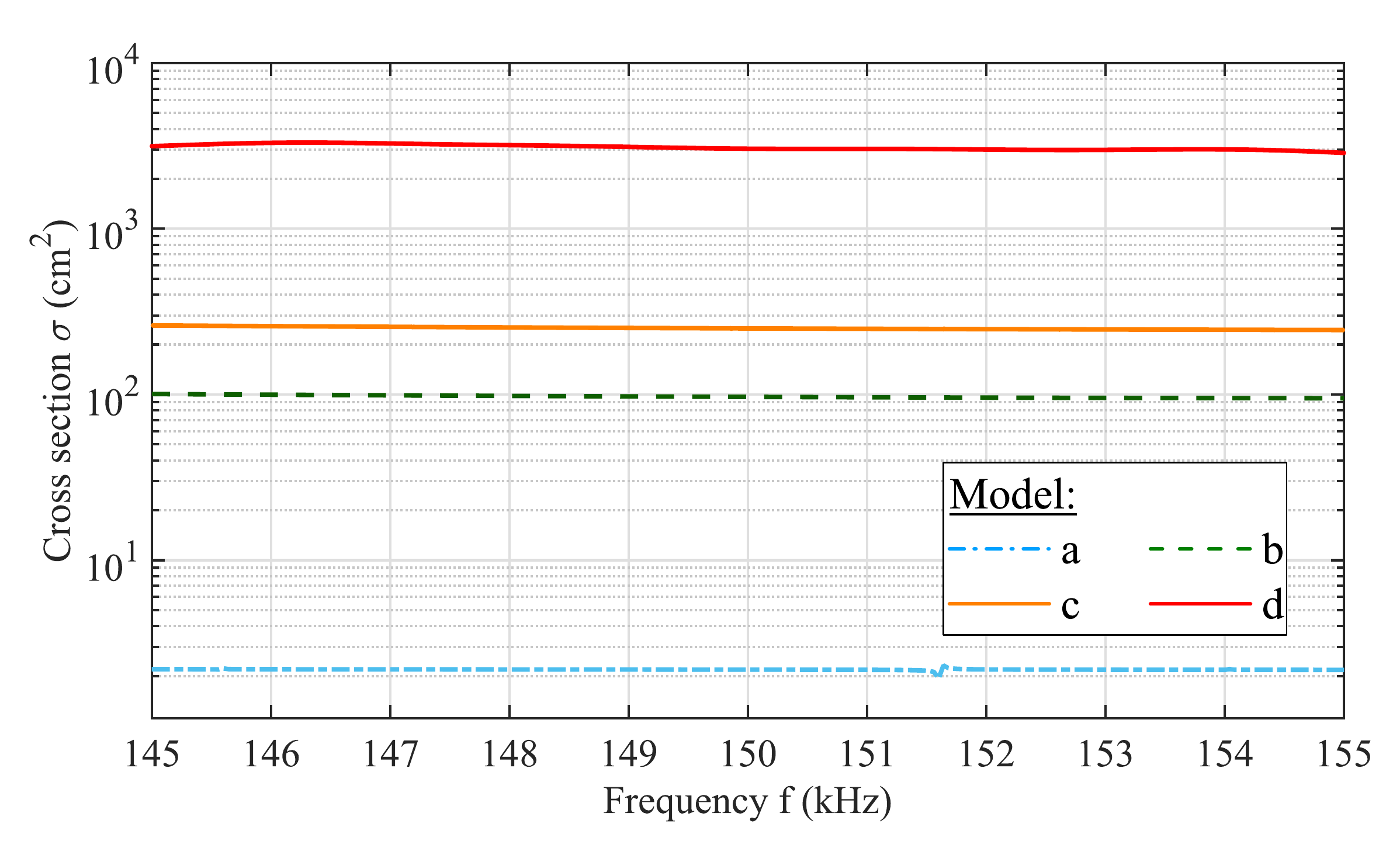}
\caption{Scattering cross section calculated for the four model scatterers.}
\label{fA1}
\end{figure}

The scattering function allows the calculation of the scattering cross-section $\sigma$ to be performed:
\begin{equation}\label{eqA5}
\sigma = \red{2\pi} \int{d\theta \sin \theta \left| F(\theta) \right|^2}.
\end{equation}
The scattering cross-section of our four spherical scatterers \red{(models a-d)} is represented \red{in} figure \ref{fA1}.
As expected from the energy density calculations (Fig. 4), the scattering from the simple air bubble is weak in comparison with the more elaborate spherical models b, c and d.
The  significantly larger scattering strengths for the three more complicated spherical models indicate that they are more promising candidates for achieving slower wave transport, especially model d.  However it is evident that scattering strength alone is not sufficient to explain the energy velocity predictions obtained from these models (see Fig. 3(a) of the main paper).  %the importance ofin the slow wave transport.

\begin{figure}[ht]
\centering
\includegraphics[width=0.8\linewidth]{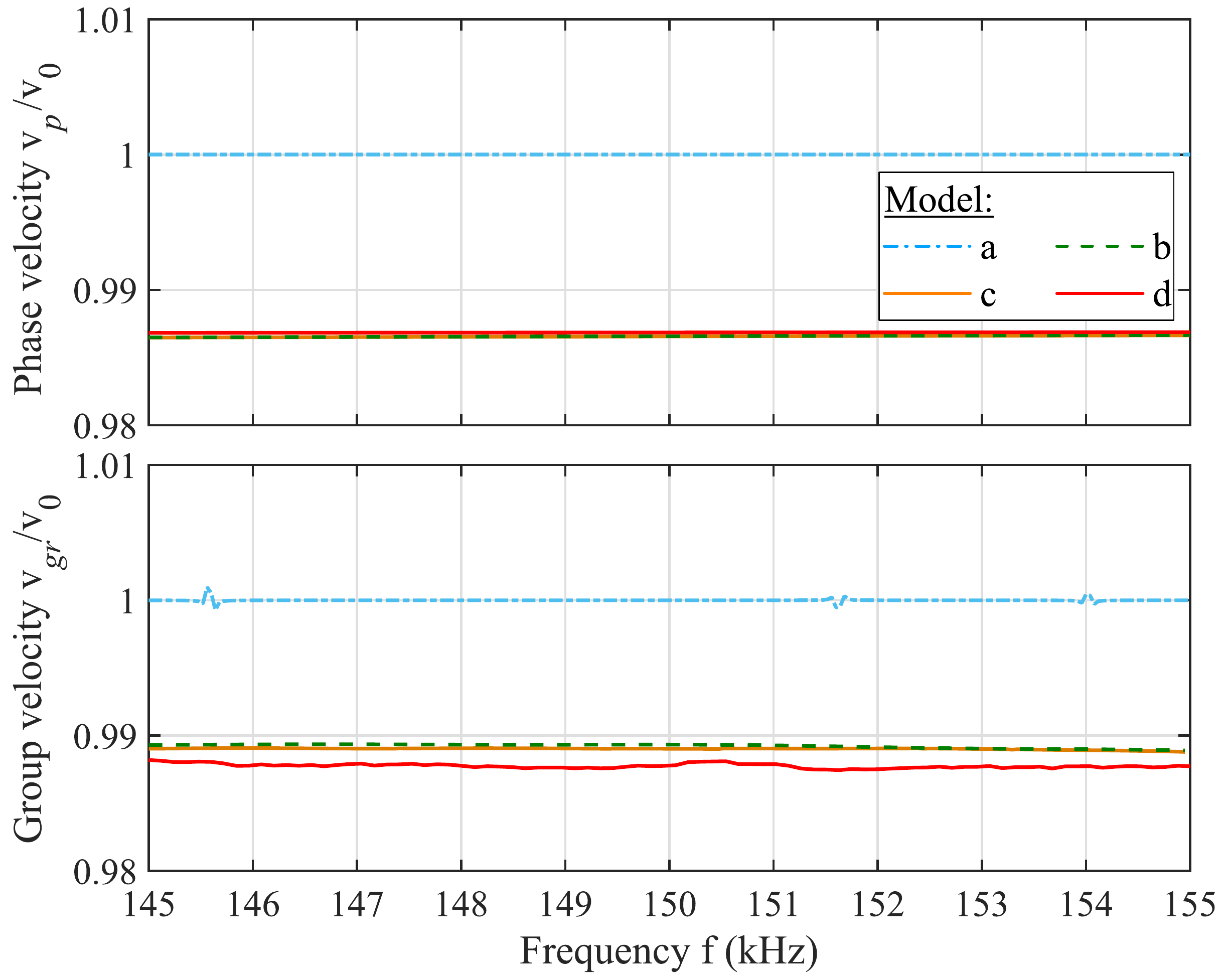}
\caption{Phase (upper panel) and group (lower panel) velocities calculated for the four different models.}
\label{fA2}
\end{figure}

The knowledge of the wave number $k$ leads to the calculation of the phase velocity $v_{p} = \omega/\textrm{Re}[k]$ and the group velocity $v_\textrm{gr} = \partial\omega/\partial\textrm{Re}[k]$.
Both these velocities are plotted in figure~\ref{fA2} for the four model scatterers.  In all cases, both of these velocities are very close to the water velocity $v_0$. 
These calculations reveal the weak phase shift of the forward scattered waves (since $v_p$ and $v_\textrm{gr}$ only depend on $F(\theta = 0)$ and $\partial\textrm{Re}[F(\theta=0)]/\partial\omega$, %its derivative with frequency
respectively).
In terms of energy transport, these results indicate that the transport velocity of the average wave field ($v_\textrm{gr}$) is barely affected by the scattering.  %therefore very close to the sound speed in pure water.
Thus, the ultra-slow transport of diffusive waves \red{must result} only %results 
from the very large scattering delay \red{of multiply scattered waves} that is described in the next section.

\section*{Energy velocity}

The energy velocity calculation is based on the renormalization of the sound speed in pure water \red{due to the scattering delay}\cite{vanAlbada91,Tallon20}
\begin{equation}\label{eqA6}
v_\textrm{e} =  \frac{v_\textrm{0}^2/v_\textrm{p}}{1 + \delta},
\end{equation}
where the \red{delay} parameter $\delta$ is equal to
\begin{equation}\label{eqA7}
\delta = 2\pi\eta v_\textrm{gr}\left ( \frac{v_\textrm{p}}{\omega}\frac{\partial\textrm{Re}F(0)}{\partial\omega} + \int_0^\pi\sin(\theta)|F(\theta)|^2\frac{\partial\varphi(\theta)}{\partial\omega}\textrm{d}\theta \right ).
\end{equation}
Under the approximation $v_0^2/v_pv_{gr}\approx1$ (for weak dispersion of the average wave field), the energy velocity can be rewritten as\cite{Schriemer97}
\begin{equation}\label{eqA8}
\frac{1}{v_e} = \frac{1}{v_{gr}} + \eta\sigma\Delta t_{ave},
\end{equation}
where the average scattering delay $\Delta t_{ave}$ is equal to
\begin{equation}\label{eqA9}
\Delta t_{ave} = \frac{\int{d\theta \sin \theta \left| F( \theta ) \right|^2}\frac{\partial \varphi(\theta)} {\partial \omega} }{\int{d\theta \sin \theta \left| F( \theta) \right|^2}} =2\pi \int{d\theta \sin \theta \Delta t(\theta)}.
\end{equation}
The right side of this equation follows from the definition of the total scattering cross section \red{[equation (\ref{eqA5})]}, %$\sigma=2\pi \int{d\theta \sin \theta \left| F( \theta) \right|^2} $, 
and is expressed in terms of the angle-dependent scattering delay $\Delta t(\theta) =  \left| F( \theta ) \right|^2\frac{\partial \varphi(\theta)} {\partial \omega}/\sigma$.   While the \red{average-field velocities only involve the scattering function in the forward direction, the scattering delay of diffusive waves %is here expressed as a function of 
depends on the frequency derivative of phase shift in \emph{all} scattering directions} $\partial\varphi(\theta)/\partial\omega$.
The group velocity being similar to the sound speed in pure water ($v_{gr}\sim v_0$), \red{it is this scattering delay that is responsible for the remarkably slow values of the energy velocity obtained for the multi-layer sphere (model d).} %this scattering delay is responsible for the renormalization of energy velocity obtained with the multi-layer sphere. This remark is particularly true for model d for which we obtained the very low $v_e$ value.
Thus, the adding of the hard scales layer significantly slows down the \red{diffusive wave transport by facilitating the storing of acoustic energy within the scatterer, and the releasing of} this energy with a large delay.